\documentclass[10pt,letterpaper,twocolumn]{article} 

\usepackage{ol2}
\usepackage[draft]{hyperref}
\usepackage{amsmath}
\usepackage{braket}
\usepackage{SIunits}
\usepackage{booktabs}
\usepackage{color}

\begin{document}

\twocolumn[ 
\title{Testing spectral filters as Gaussian quantum optical channels}

\author{K.~Laiho,$^{1}$ A.~Christ,$^{1,2}$ K.~N.~Cassemiro$^{1}$ 
and C.~Silberhorn$^{1,2}$}

\address{$^{1}$ Max Planck Institute for the Science of Light, G\"unther-Scharowsky-Stra\ss e 1/Bldg.~24, D-91058 Erlangen, Germany \\
$^{2}$ Applied Physics, University of Paderborn,  Warburgerstra\ss e 100, D-33098 Paderborn, Germany}


\begin{abstract}
We experimentally investigate the mode characteristics of multimode radiation fields propagating through  frequency dependent Gaussian channels. After manipulating the twin beams emitted from a conventional parametric down-conversion source via spectral filtering, we study the changes in their  mode characteristics, utilizing the joint  normalized correlation functions.  While filtering reduces  the number of spectral modes, it  also leads to an apparent mode mismatch, which destroys the perfect photon-number correlation between the twin beams, and  influences the mode properties of  heralded states.
\end{abstract}

 ] 

Quantum channels are basic building blocks  of many quantum optical applications, in particular of quantum communication protocols. Often, these communication channels, such as optical fibers, are Gaussian channels, which introduce losses and thus lead to decoherence  limiting the channel performance~\cite{Barbosa2010}. This becomes a more prominent problem if channel multiplexing is taken into account, especially for pulsed light exhibiting  a broad spectrum.
Therefore, efficient and robust  methods for  characterizing quantum optical multimode (MM) light fields gain importance in the future quantum-communication systems. In the past, loss-tolerant photon counters have been used for  exploring the mode structure of quantum light~\cite{Paleari2004, Wasilevski2008, M.Avenhaus2008, W.Mauerer2009, Laiho2009 }.
Nonetheless,  determining the mode number from photon statistics without  \emph{a priori} knowledge of the mode distribution is a challenging task.

In this Letter we experimentally investigate the normalized correlation functions  (nCFs) \cite{Avenhaus2010},  which provide \emph{loss-independent} techniques for characterizing the effective numbers of the excited modes. The nCFs are closely related to the factorial moments of  the photon number and can also disclose the spatiotemporal mode properties of the quantum states~\cite{Tapster1998,  Gisin2004, Christ2010}. 
The measurement  of nCFs is typically realized
by splitting the light field in a beam splitter network,  followed by coincidence photon counting~\cite{Avenhaus2010}. 
This technique probes the complete photon-number content of the state. Thus, it has a distinct advantage over two-photon quantum interference experiments  that normally explore only the single-photon components of the interfering states  and regard the higher photon-number contributions as undesired background \cite{Mosley2008, Laiho2009, Cassemiro2010}. 
The nCFs have successfully  been employed to study the photon-number content of quantum states~\cite{Kimble1977, Bussieres2008,  Avenhaus2010}, to investigate the single-mode (SM) properties of nonclassical light fields~\cite{Tapster1998, Li2008, Blauensteiner2009, A.Eckstein2010}, and to demonstrate the temporal photon-number correlation between photon pairs~\cite{Bocquillon2009, M.Scholz2009}.

\emph{Twin beams}, produced in parametric down-conversion (PDC), exhibit strict photon-number correlation but are conventionally also  highly correlated in the spectral degree of freedom; in other words, they possess spectral entanglement \cite{W.P.Grice1997}.  This might be especially appealing for the channel multiplexing since a set of squeezed states is generated in orthogonal modes that can be transmitted simultaneously \cite{Wasilewski2006}. However, if MM radiation is used without care in the standard quantum-key-distribution protocols, they can introduce  a severe security risk~\cite{ W.Helwig2009}: the mode characteristics can be manipulated in the channel e.g.,~by spectral filtering.
Contrariwise, regarding the preparation of heralded nonclassical states of light, such as photon-number Fock states~\cite{A.I.Lvovsky2001, A.Ourjoumtsev2006},  decorrelation of the used twin beams is essential.
Therefore, spectral and spatial filtering is often involved in the state generation to ensure the SM characteristics.

Spectral filtering can be rigorously described as a frequency dependent beam splitter, which destroys  the perfect photon-number correlation between the twin beams~\cite{A.M.Branczyk2009}. Moreover, filtering disrupts the orthogonality of the previously independent modes, and it is not trivial to predict the remaining  number of modes. Here, we determine the mode properties of spectrally correlated twin beams by measuring their joint, low-order nCFs in a time-integrated form. 
After modifying the mode structure by spectral filtering, we record different effective mode numbers  for the twin beams.
This indicates that any model which regards the filtered twin-beam state simply  as a product of pure two-mode squeezers  with reduced number of modes incorrectly assumes equal mode numbers for the conjugate beams.
Additionally, we investigate the mode characteristics of heralded quantum states of light.

 The spectral structure of unfiltered twin beams, denoted as signal ($s$) and idler  ($i$), 
  is described by the joint correlation function $f(\nu_{s}, \nu_{i})$ in terms of the frequencies $\nu_{\mu}$ ($\mu = s, i$). The properties of   $f(\nu_{s}, \nu_{i})$ are defined by the energy and momentum conservation laws~\cite{W.P.Grice1997}.  The former is determined by the  spectral shape of the pump field  and the latter by the dispersion of the nonlinear medium, resulting in the so-called phase-matching function. According to our earlier studies \cite{Laiho2009},  the employed source---a  \unit{1.45}{\milli \meter} long periodically poled type-II KTP waveguide (WG)---produces highly spectrally anti-correlated twin beams. Because of the type-II process, the signal and idler marginals also exhibit different spectral widths, and thus filtering has different impacts on them.

  

 The broadband mode structure of pure twin-beam states is accessible  by means of a singular-value decomposition, $f(\nu_{s}, \nu_{i}) = \sum_{k} \lambda_{k} \phi_{k}(\nu_{s})\psi_{k}(\nu_{i})$, where $\{  \phi_{k}(\nu_{s})\}$ and $\{\psi_{k}(\nu_{i})\}$ each form a set of orthonormal functions \cite{Law2000}.  The normalization is chosen such that  $\sum_{k}{\lambda_{k}^{2}}= 1$. This decomposition is unique, and it guarantees  a perfect  mode correlation between  signal and idler. The number of excited modes can be determined  with the K-parameter $K= 1/\sum_{k}{\lambda_{k}^{4}}$, which indicates  the effective number of  uniformly occupied modes \cite{Eberly2005}. 
We gain information of this mode structure by exploring the joint nCFs~\cite{Avenhaus2010}.  In our time-integrated measurements, they probe a superposition of all modes and provide information of the complete mode structure~\cite{Christ2010}: $g^{(n,m)} = \frac{ \braket{ : (\sum_k \hat{A}_k^{\dagger}    \hat{A}_k) ^{n}( \sum_{k'}\hat{B}_{k'}^{\dagger} \hat{B}_{k'})^{m}: } }{ \braket{ \sum_k \hat{A}_k^{\dagger}    \hat{A}_k } ^{n} \braket{ \sum_{k'} \hat{B}_{k'}^{\dagger } \hat{B}_{k'} } ^{m}}$. The operators $\hat{A}_k^{\dagger}(\hat{A}_k$) and  $ \hat{B}_k^{\dagger} ( \hat{B}_k)$  describe  the creation (annihilation) of a photon in the  $k$th broadband signal and idler modes.

\begin{figure}
\includegraphics[width = 0.48\textwidth]{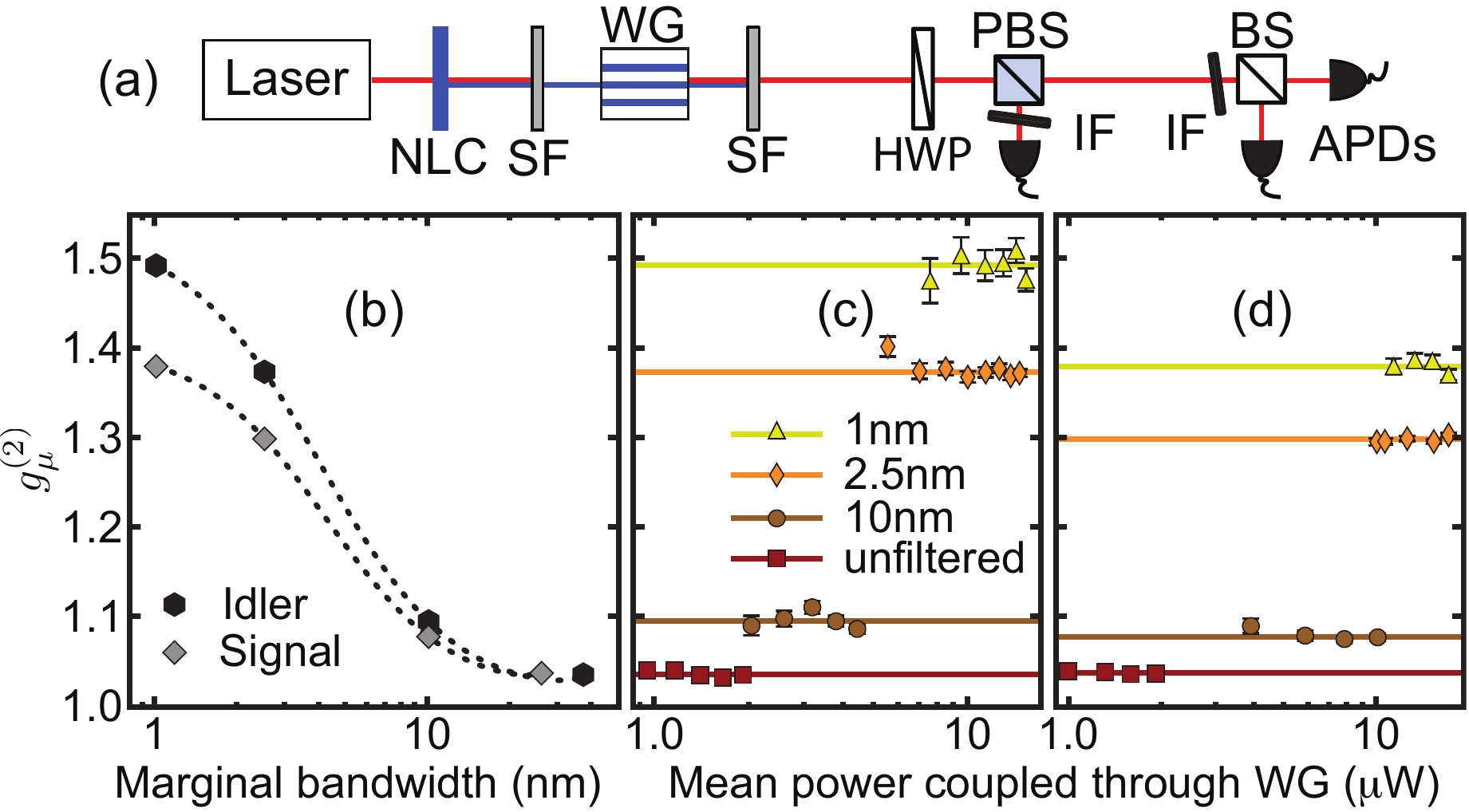}
\caption{\label{Fig_g_2_0} (a) Experimental setup. (b) $g_{\mu} ^{(2)}$ vs.~the filtered marginal bandwidth. (c)  $g_{i}^{(2)}$ vs.~the pump power. (d) Same as (c) for $g_{s}^{(2)}$.  Symbols are measured values, solid lines fits, and dotted lines guides for the eye.}
\end{figure}

In the experiment, illustrated in Fig.~\ref{Fig_g_2_0}(a), Ti:Sapphire laser pulses (\unit{796}{\nano\meter}, \unit{10}{\nano\meter} bandwidth, \unit{4}{\mega\hertz} repetition rate) were frequency doubled in a nonlinear crystal (NLC) and employed as a pump for the PDC process.  The residual pump fields were blocked with spectral filters (SF) and  signal and idler were separated in a polarizing beam splitter (PBS). One of the twin beams, selected by the setting of a half-wave plate (HWP), was directed to a \unit{1}{\nano\meter} broad interference filter (IF) and then detected with an avalanche photodiode (APD). The other one was optionally filtered to a desired bandwidth of \unit{1}{\nano\meter}, \unit{2.5}{\nano\meter} or \unit{10}{\nano\meter} with exchangeable spectral filters, and then split by a symmetric beam splitter (BS), and launched  to two APD detection arms.
 The spatial SM characteristics of the twin beams were ensured by utilizing SM fibers.
We recorded the coincidence and single counts for the evaluation of the nCFs from the measured raw counts according to \cite{Avenhaus2010}. Furthermore, nanosecond time gatings were used in all the detection channels to suppress background and dark count events. 

First, we evaluate the second order moments $g^{(2,0)}$ and $g^{(0,2)}$, shortly $ g_{\mu}^{(2)}$. They obey  $1< g_{\mu}^{(2)} \le 2$,  with the  boundaries indicating the MM (\emph{Poissonian}) and SM (\emph{thermal}) behavior of the marginal beams \cite{M.Avenhaus2008}.  The measured values are shown in Fig.~\ref{Fig_g_2_0}(b) with respect to the bandwidths of the marginal beams, estimated by the filter bandwidths or the spectral widths of the unfiltered marginal distributions. Assuming a loss-free PDC generation process,
we can extract the effective mode number in each twin beam without \emph{a priori} information of the weights of the mode distribution \cite{Christ2010}: in the low power regime $g_{\mu}^{(2)}$ is \emph{independent} of the pump power and related to the corresponding K-parameter 
by $g_{\mu}^{(2)} = 1+1/K_{\mu}$. 
 By repeating the measurement at different pump powers we confirm that  the value of $g_{\mu}^{(2)}$ is unchanged as shown in Figs~\ref{Fig_g_2_0}(c-d). 
The K-parameters extracted from the measurements are shown in  Table \ref{Table_K} with an accuracy given by the statistical fluctuations of $g_{\mu}^{(2)}$.
 As expected, the unfiltered marginal beams exhibit a large number of modes. Moreover,  the mode numbers of the unfiltered twin beams coincide due to the perfect photon-number correlation of  pure two-mode squeezed states. Spectral filtering leads to different effective mode numbers in signal and idler, which can only be explained by the fact that the perfect photon-number correlation between them is destroyed. Overall, $g_{\mu}^{(2)}$  provides a sensitive measure of the mode number mismatch and directly indicates that 
the filtered state is not perfectly two-mode squeezed anymore.

\begin{table}[!b]
\begin{center}
\begin{tabular}{ccccc}
\toprule
 filter (\unit{\nano\meter}{} \hspace{-1.5ex}) &  1 &2.5 &10 & $\infty$\\
\midrule
 $K_{s}$  & 2.63(3) & 3.33(2) & 12.7(4) & 26(2) \\
 $K_{i}$     & 2.03(3)   &2.67(2)     & 10.4(4)    & 27.3(8) \\
\bottomrule
\end{tabular}
\caption{\label{Table_K}The K-parameters of signal ($K_{s}$) and idler ($K_{i}$). }%
\end{center}
\end{table}

\begin{figure}
\includegraphics[width = 0.48\textwidth]{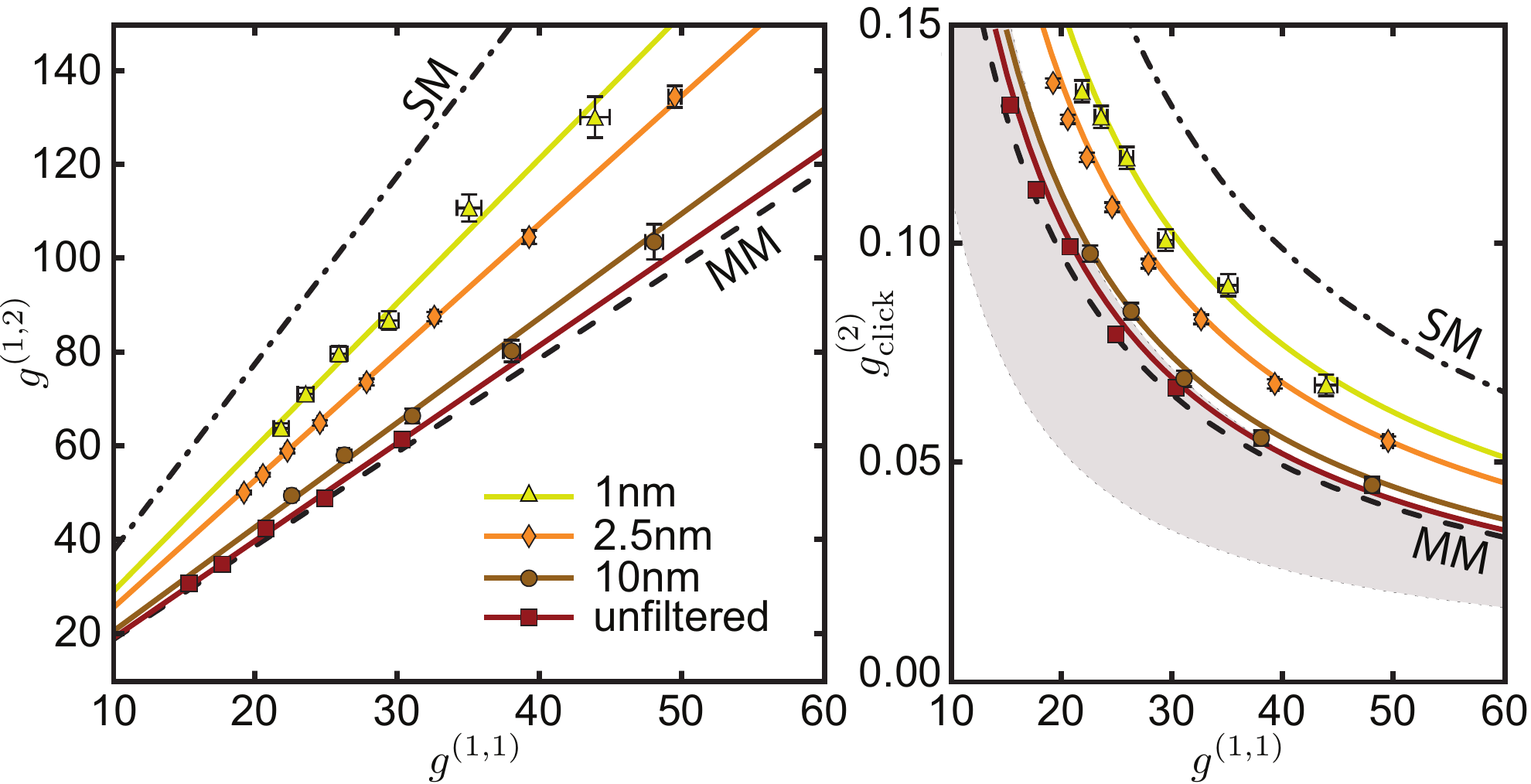}
\caption{\label{Fig_g_2_1} (a) $g^{(1, 2)}$  vs.~$g^{(1,1)}$. Solid lines are linear fits. (b) $g^{(2)}_{\textrm{click}}$  vs.~$g^{(1,1)}$. Solid lines are predicted by the fits in (a). Symbols indicate measured values, and dashed (dash-dotted) lines illustrate the MM (SM) boundaries. The shaded area shows the boundaries for lossless detection.}
\end{figure}

Next, we examine the joint, low-order nCFs  $g^{(1,2)}$ and $g^{(1,1)}$.  For a pure twin-beam state, the SM  and MM characteristics are predicted by   $g^{(1,2)}_{SM} = 4 g^{(1,1)} -2$ and $g^{(1,2)}_{MM} = 2 g^{(1,1)} -1$. 
Our results, illustrated in Fig.~\ref{Fig_g_2_1}(a), lie in between these boundaries. Furthermore, one clearly recognizes the linear behavior of  the curves, whose slope is ideally governed by $ 2g_{\mu}^{(2)}$. The linear fits are in a good agreement with  the measured values of $ g_{i}^{(2)}$. At higher gains several modes can be excited simultaneously and tight spectral filtering employed in only one of the twin beams is not sufficient to ensure the SM behavior of the joint state [Fig.~\ref{Fig_g_2_1}(a)-squares]. The mode structure is  modified  if filtering in both arms is introduced [Fig.~\ref{Fig_g_2_1}(a)-triangles]. Intriguingly, these results of the filtered twin-beam states are similar to those expected for  photon-number correlated twin-beam states. Therefore, the loss-independent nCFs are partly insensitive to the disappearance of the strict photon-number correlation.
 
Regarding the preparation of  heralded single photons we investigate the conditional  second order moment $g^{(2)}_{\textrm{click}}$ of idler, triggered by a click in signal. 
In  Fig.~\ref{Fig_g_2_1}(b) we present $g^{(2)}_{\textrm{click}}$   with respect to $g^{(1,1)}$.
A decrease in the pump power increases $g^{(1,1)}$, which indicates a high photon-number correlation between signal and idler, a  desirable feature for heralding \cite{Christ2010}. Nevertheless, the complete suppression of the higher photon-number contributions in the heralded state is experimentally challenging and happens at the cost of the source brightness.
Unlike the joint nCFs,   $g^{(2)}_{\textrm{click}}$ depends on the  efficiency in the trigger arm. According to our earlier studies \cite{Laiho2009},
 the second conditional moment is  expressed in the low efficiency regime as $g^{(2)}_{\textrm{click}}= g^{(1,2)}/[g^{(1,1)}]^{2}$ for photon-number correlated twin beams. We predict the behavior of  $g^{(2)}_{\textrm{click}}$  with the help of the linear fits in Fig.~\ref{Fig_g_2_1}(a)  and find a good agreement with the measurement.  This indicates that the MM structure is still visible in the higher photon-number contributions of the heralded state. Moreover, the values of  $g^{(2)}_{\textrm{click}}$, as illustrated in Fig.~\ref{Fig_g_2_1}(b),  are significantly larger than expected for a lossless detection in the trigger arm. 

We studied the changes in the mode structure of twin beams by investigating their joint nCFs.
We quantified the effective mode number in the twin beams with the K-parameter after modifying their spectral properties  in frequency-dependent Gaussian channels. Our results show that the perfect photon-number correlation between signal and idler is lost. 
 Furthermore, we gain a deeper insight into the spectral mode structure of the studied states, although  the nCFs cannot resolve the weights of the individual modes.  Our results can have a significant impact on controlling the mode properties by spectral filtering and will become important for quantum communication protocols implemented in optical fibers.

We thank M.~Avenhaus  and A.~Schreiber for helpful discussions and useful comments.
This work was supported by the EC under the grant agreement CORNER (FP7-ICT-213681). KNC~acknowledges  the financial support from the Alexander von Humboldt  Foundation.

\end{document}